\documentclass[
reprint,
nofootinbib,
amsmath,amssymb,
aps,
prl,
]{revtex4-1}
\usepackage{graphicx}
\usepackage{dcolumn}
\usepackage{bm}
\usepackage{color}
\usepackage[T1]{fontenc}
\usepackage{mathrsfs,amsfonts,dsfont}
\usepackage{amstext}
\usepackage{mathtools}
\usepackage{enumitem}
\usepackage{changes}
\usepackage{longtable}
\newcommand{\tr}[1]{\text{tr}\left(#1\right)}
\newcommand{\ket}[1]{\left\vert#1\right\rangle}
\newcommand{\bra}[1]{\left\langle#1\right\vert}
\newcommand{\gobs}{g_\text{obs}}
\newcommand{\pguess}{P_\text{g}}
\begin{document}
\preprint{APS/123-QED}
\title{A Genuine Multipartite Bell Inequality for \\ Device-independent Conference Key Agreement}
\author{Timo Holz}\email{holzt@uni-duesseldorf.de}
\author{Hermann Kampermann}
\author{Dagmar Bru\ss}
\affiliation{Institut f\"ur Theoretische Physik $\MakeUppercase{\romannumeral3}$, Heinrich-Heine-Universit\"at D\"usseldorf, D-40225 D\"usseldorf, Germany}
\date{\today}
\begin{abstract}
In this work, we present a new class of genuine multipartite Bell inequalities, that is particularly designed for multipartite device-independent (DI) 
quantum key distribution (QKD), also called DI conference key agreement. 
We prove the classical bounds of this inequality, discuss how to maximally violate it and show its usefulness by 
calculating achievable conference key rates via the violation of this Bell inequality. To this end, semidefinite programming techniques 
based on [Nat. Commun. 2, 238 (2011)] are employed and extended to the multipartite scenario. 
Our Bell inequality represents a nontrivial multipartite generalization of the Clauser-Horne-Shimony-Holt inequality and is motivated by the extension 
of the bipartite Bell state to the $n$-partite Greenberger-Horne-Zeilinger state. 
For DIQKD, we suggest an honest implementation for any number of parties and study the effect of noise on achievable asymptotic conference key rates.
\end{abstract}
\maketitle
\emph{Introduction.---} 
Among a variety of quantum technology applications~\cite{motivation1,motivation2,motivation3}, quantum key distribution (QKD) 
is one of the most prominent concepts, in particular for multiple parties in a quantum network~\cite{NQKD}. 
Early proposed QKD protocols~\cite{BB84,ekert,6state} have high demands on experimental assumptions which are difficult to guarantee. 
Device-independent (DI) QKD aims at establishing a secret key without making detailed assumptions about the inner 
working processes of the quantum devices~\cite{DI1, DI2005, DIcolbeck, scaraniDI, scaraniDIlong}.\\
The security of DIQKD protocols is based on a loophole-free violation of a Bell 
inequality~\cite{scaraniDI,scaraniDIlong,Masanes,secproof,fullydiqkd,rotem1,rotem2,DICKAfixed}. 
A connection between the DI secret-key rate and the violation of the associated Clauser-Horne-Shimony-Holt (CHSH) inequality~\cite{chsh} was 
established in~\cite{scaraniDI,scaraniDIlong} for the bipartite setting. 
In Ref.~\cite{DICKAfixed}, a protocol to generate a secret key among $n$ parties, called DI conference key agreement (DICKA) 
was introduced, which relies on the violation of the Parity-CHSH inequality. Hereby, nonlocality is certified via an effective  
Bell test of two parties depending on the measurement results of the remaining ones.

\begin{figure}[ht!]
\centering
\includegraphics[width=0.39\textwidth]{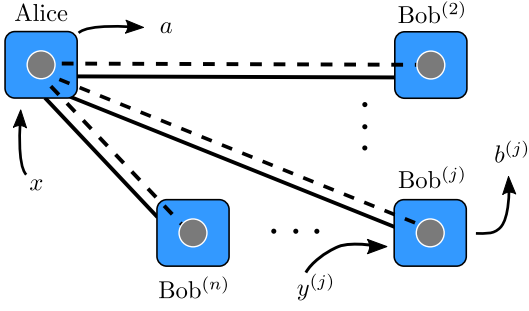}
\caption{A multipartite DIQKD setting, with parties Alice and $\big\{$Bob$^{(j)}\big\}_{j=2}^{n}$. Alice distributes a multipartite quantum state 
via quantum channels (dashed lines). The parties communicate over classical channels (solid lines) and they perform measurements 
on their part of the quantum resource, specified via an input $x, y^{(j)}\in\{0,1\}$ that yields a result $a, b^{(j)}\in\{\pm1\}$.}
\label{fig:nmksetting}
\end{figure}

Not all multipartite Bell inequalities are suitable for DIQKD because measurements and quantum resources are 
required that allow a sufficiently large Bell-inequality violation and at the same time provide highly correlated measurement results among all 
parties. Moreover, at least one party has to use one measurement for key generation \emph{and} for the Bell test, to detect a potential tampering 
of the devices. Achieving these requirements simultaneously should therefore be guaranteed by the very structure of the Bell inequality. This 
constraint disqualifies several known Bell inequalities as a viable option for a Bell test in DIQKD with certain quantum states. For instance, the 
archetypical $n$-partite Greenberger-Horne-Zeilinger (GHZ) state~\cite{ghz} 
can maximally violate the $n$-partite Mermin-Ardehali-Belinski{\u\i}-Klyshko (MABK) inequality~\cite{mermin, ardehali, belinskii} and also the 
Bell inequality most recently introduced in Ref.~\cite{tailoredBell}. However, as proven in Ref.~\cite{NQKD}, perfectly correlated measurement results 
with the $n$-GHZ state can only be obtained if and only if all parties measure in the $\sigma_z$ eigenbasis, which then excludes maximum violation of 
the Bell inequalities in Refs.~\cite{mermin,ardehali,belinskii,tailoredBell}, see~\cite{comment}.\\
In this work, we specifically design a novel class of multipartite Bell inequalities that fulfills the aforementioned conditions. 
We prove the classical bounds of 
this inequality and discuss some features of it, in particular how to obtain a large Bell-inequality violation. 
To demonstrate the usefulness of our Bell inequality, we quantify achievable conference key rates based on its violation. 
For this, we use the approach of Ref.~\cite{Masanes}, which employs the Navasqu\'{e}s-Pironio-Acin (NPA) hierarchy~\cite{npa0,npa}, together with 
a multipartite constraint. We propose an honest implementation for a multipartite DIQKD protocol and briefly discuss how noise affects the 
achievable asymptotic DI secret conference key rates.\\ 
\emph{A genuine multipartite Bell inequality.---}
We impose the following condition on the Bell test: 
Its structure has to be such that it allows to simultaneously yield highly correlated measurement results 
and sufficiently large Bell-inequality violation for certain quantum states. These are crucial ingredients in any DIQKD protocol.\\
Consider a setup of $n$ parties, called Alice and Bob$^{(j)}$ for $j\in\{2,\dots,n\}\eqqcolon[n]$, cf. Fig.~\ref{fig:nmksetting}.
Let each party measure two dichotomic observables $A_x$ and $B_{y^{(j)}}^{(j)}$, with inputs $x,y^{(j)}\in\{0,1\}$. 
We define a set that contains all ordered possibilities to choose $l$ out of the labels $\{2,\dots,n\}$ for the Bobs:
\begin{align}\label{set}
\mathcal{S}^{(n)}_l\coloneqq \Big\{
\boldsymbol{\alpha}_l^{(n)}\coloneqq\Big(\alpha^{(n)}_{l,1},&\dots,\alpha^{(n)}_{l,l}\Big) \Big\vert \,\, \alpha^{(n)}_{l,j} < \alpha^{(n)}_{l, j+1} \\
&\forall j\in\{1,\dots,l-1\}, \alpha_{l,j}^{(n)} \in [n]\Big\},  \nonumber 
\end{align}
for all $n\in\mathbb{N}, l \in \{1,\dots,n-1\}$, with vectors $\boldsymbol{\alpha}_l^{(n)}$ of length $l$, whose ordered components $\alpha_{l,j}^{(n)}$ 
label a specific Bob; e.g., $\mathcal{S}_2^{(4)}=\{(2,3),(2,4),(3,4)\}$.
For the sake of legibility, we also use the abbreviation 
\begin{align}
 B_\pm^{(j)}\coloneqq \frac{1}{2}\big(B_0^{(j)}\pm B_1^{(j)}\big). \label{Bpm}
\end{align}

\emph{Definition. (Genuine multipartite Bell inequality)} 
Let $n\geqslant3$ be an integer and $\mathcal{S}^{(n)}_l$ the set defined in Eq.~\eqref{set}. 
\begin{align}\label{BellIneqClass}
\mathcal{B}^{(n)}&\coloneqq\Bigg\langle A_1 \bigotimes\limits_{j=2}^{n}B_+^{(j)}\Bigg\rangle 
- \delta_{\lfloor\frac{n}{2}\rfloor,\frac{n}{2}} \left\langle A_0 \bigotimes\limits_{j=2}^{n}B_-^{(j)}\right\rangle \\ \nonumber 
&-\sum\limits_{k=1}^{\lfloor\frac{n-1}{2}\rfloor}\Bigg[
\left\langle 
A_0\otimes \sum\limits_{\boldsymbol{\alpha}^{(n)}_{2k-1}\in\mathcal{S}^{(n)}_{2k-1}}\bigotimes\limits_{j=1}^{2k-1} B_-^{\left(\alpha^{(n)}_{2k-1,j}\right)}
\right\rangle \\ \nonumber 
&{}\qquad\quad+\left\langle 
\sum\limits_{\boldsymbol{\alpha}^{(n)}_{2k}\in\mathcal{S}^{(n)}_{2k}}\bigotimes\limits_{j=1}^{2k} B_-^{\left(\alpha^{(n)}_{2k,j}\right)}
\right\rangle 
\Bigg] 
\quad
\begin{cases}
\begin{alignedat}{1}
&\leqslant g_\text{cl}^{(n)\downarrow} \\
&\geqslant g_\text{cl}^{(n)\uparrow}
\end{alignedat}
\end{cases}
\end{align}
defines a genuine multipartite Bell inequality, with upper and lower classical bound $g_\text{cl}^{(n)\downarrow}$ and 
$g_\text{cl}^{(n)\uparrow}$, respectively.

\begin{figure}[ht!]
\centering
\includegraphics[width=0.39\textwidth]{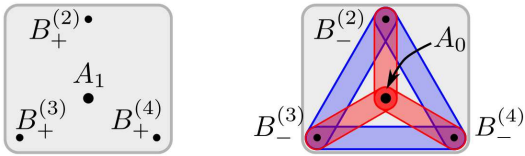}
\caption{Graphical representation of the correlators in the Bell inequality~\eqref{BellIneqClass} for $n=4$, which 
highlights the special role of Alice and the symmetry of the inequality w.r.t. to the Bobs. Vertices denote observables, and each hyperedge symbolizes 
a correlator the contains the corresponding observables.}
\label{fig:bellcorr}
\end{figure}

Remember that $B_{+}^{(j)}$ and $B_{-}^{(j)}$ depend on each other, see Eq.~\eqref{Bpm}. In the Suppl. Mat., we elaborate in detail on the construction of 
the Bell inequality. To make it more accessible, we state the Bell correlator for $n=3$,
\begin{align}\label{bell3}
\mathcal{B}^{(3)}=\left\langle A_1B_+^{(2)}B_+^{(3)}\right\rangle &- \left\langle A_0\!\left(B_-^{(2)}+B_-^{(3)}\right)\right\rangle \nonumber \\  
&-\left\langle B_-^{(2)}B_-^{(3)}\right\rangle,
\end{align}
and visualize it in Fig.~\ref{fig:bellcorr} for $n=4$.

\emph{Lemma. (Reduction of party number)} 
For all $n\geqslant2$, $\mathcal{B}^{(n-1)}$ is recovered from $\mathcal{B}^{(n)}$ via $B_0^{(n)}=B_1^{(n)}=\mathds{1}$.

\emph{Proof.} We have $B_-^{(n)}=0$, hence 
\begin{align}
\bigotimes\limits_{j=1}^lB_-^{\left(\alpha_{l,j}^{(n)}\right)}=0  
\quad \forall\,\, \boldsymbol{\alpha}_l^{(n)}\in\mathcal{S}^{(n)}_l\setminus\mathcal{S}^{(n-1)}_l. 
\end{align}
Therefore, the sum over the set $\mathcal{S}^{(n)}_l$ is converted into a sum 
over $\mathcal{S}^{(n-1)}_l$. For $n$ odd, the term $\langle A_0\bigotimes_{j=2}^{n-1}B_-^{(j)}\rangle$ emerges from the sum in 
inequality~\eqref{BellIneqClass} for $k = \frac{n-1}{2}$. As $B_+^{(n)}=\mathds{1}$, the proof is complete. \hfill $\blacksquare$

By iteration, $\mathcal{B}^{(k)}$ is obtained from $\mathcal{B}^{(n)}$ for all $k<n$.

\emph{Theorem. (Classical Bounds)}
In any classical theory, the lower and upper bounds on $\mathcal{B}^{(n)}$ are given by 
\begin{align}
g_\text{cl}^{(n)\uparrow} = -\left(2^{n-1}-1\right) \quad \text{and} \quad g_\text{cl}^{(n)\downarrow}= 1 \quad \forall n\in\mathbb{N}. \label{clbounds}
\end{align}

\begin{figure}[ht!]
\centering
\includegraphics[width=0.475\textwidth]{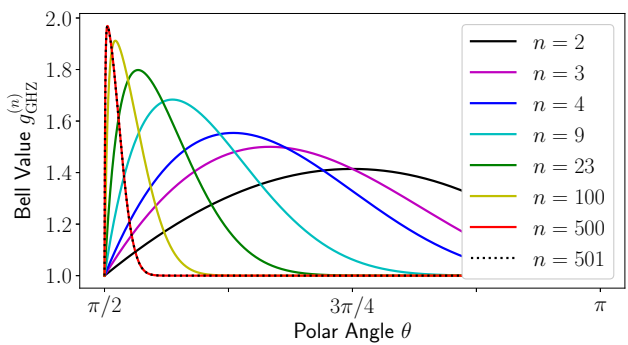}
\caption{Achievable Bell value $g_\text{GHZ}^{(n)}$ according to Eq.~\eqref{GHZvalue_simplified} as a function of the polar angle $\theta$ 
for various number of parties $n$.}
\label{fig:bellvalue}
\end{figure}

Note that the upper bound is independent of $n$. See Suppl. Mat. for the analytical proof, whose idea is to consider all classical deterministic 
strategies, which can be significantly reduced by exploiting the invariance of $\mathcal{B}^{(n)}$ under arbitrary relabeling of Bobs.\\
Here, some remarks are due. First, note that for $n=2$, $\mathcal{B}^{(n)}$ and the classical bounds reproduce the CHSH inequality (normalized with a 
factor $\frac{1}{2}$). Furthermore, the Parity-CHSH inequality~\cite{DICKAfixed} is in fact a subclass of our Bell inequality, that is recovered via the choice 
$B_0^{(j)}=B_1^{(j)}\eqqcolon B^{(j)}$ for all $j \geqslant 3$. Also, note that the lower classical bound 
on $\mathcal{B}^{(n)}$ is close to the algebraic minimum of $-2^{n-1}$. As 
we did not find a way to violate the lower bound, a \emph{violation} of the Bell inequality~\eqref{BellIneqClass} refers to the upper bound throughout 
this paper. 
Beyond that, a characterization of the maximum Bell value achievable with quantum correlations, the Tsirelson bound $g_\text{qm}^{(n)}$~\cite{tsirelson}, 
is desirable. However, there is no general approach known that yields a tight Tsirelson bound for an arbitrary Bell inequality, as mentioned in 
Ref.~\cite{tailoredBellBipartite}. An upper bound on the Tsirelson bound can be found by using the NPA hierarchy~\cite{npa}. Usually, this procedure 
is numerically expensive, which is why we only calculate this bound for the first nontrivial odd- and even-numbered case, i.e., for $n\in\{3,4\}$:
\begin{align}
 g_\text{qm}^{(3)} = 1.5 \quad \text{and} \quad g_\text{qm}^{(4)} \approx 1.5539. \label{tsirelson34}
\end{align}
These bounds are tight within numerical precision, cf. Table~\ref{table:table}. The Bell inequality~\eqref{BellIneqClass} is particularly 
designed for the state $\ket{\text{GHZ}_n}=\frac{1}{\sqrt{2}}\big(\ket{0}^{\otimes n} + \ket{1}^{\otimes n}\big)$, under the condition that 
the choice $A_0=\sigma_z$ does not prohibit a violation of this inequality. The optimal measurements can be chosen to be in the 
$\sigma_z-\sigma_x$ plane of the Bloch sphere, as further argued in the Suppl. Mat., in detail,
\begin{subequations}
\label{optimalchoice}
\begin{align}
A_0 &= \sigma_z,  &B_0^{(j)}&= \sin(\theta)\sigma_x + \cos(\theta)\sigma_z, \\ 
A_1 &= \sigma_x,  &B_1^{(j)}&= \sin(\theta)\sigma_x - \cos(\theta)\sigma_z, 
\end{align}
\end{subequations}
for all $j\in[n]$, where the optimal value of the polar angle $\theta$ depends on the number of parties $n$. Note that, 
due to the symmetry of the Bell correlator and 
the target state, $\theta$ does not depend on $j$. 
This choice allows a straightforward calculation of the Bell value achievable with the $n$-GHZ state, which reads
\begin{subequations}
\label{GHZvalue_simplified}
\begin{align}
g^{(n,\text{odd})}_{\text{GHZ}}\!\!&= 1- \left(1+\cos\left(\theta\right)\right)^{n-1}\!+ \sin^{n-1}\!\left(\theta\right), \\ 
g^{(n,\text{even})}_{\text{GHZ}}\!\!&= 1- \left(1+\cos\left(\theta\right)\right)^{n-1}\!+ 
\frac{\cot\left(\theta\slash2\right)\sin^n\!\left(\theta\right)}{1+\cos\left(\theta\right)}.
\end{align}
\end{subequations} 
Table~\ref{table:table} displays some quantities of interest for $n\leqslant 7$. 
\renewcommand{\arraystretch}{1.25}
\setlength{\LTcapwidth}{8.5cm} 
\begin{longtable}[]{p{1.4cm} p{2.2cm} p{2.2cm} p{2.2cm} }
\caption{Maximum Bell value $g_\text{GHZ}^{(n)}$ achievable with $n$-GHZ state, cf. Eq.~\eqref{GHZvalue_simplified}, the ratio of 
$g_\text{GHZ}^{(n)}$ and $g_\text{GHZ}^{(n-1)}$, and the corresponding polar angle $\theta$ for all Bobs. 
The quantum-to-classical ratio is given by $g_\text{GHZ}^{(n)}$, as $g_\text{cl}^{(n)\downarrow}=1$ for all 	$n$. 
The values are rounded to the fourth decimal place.}\\
	
$\mathcal{B}^{(n)}$ & $g_\text{GHZ}^{(n)}$		& $g_\text{GHZ}^{(n)}\slash g_\text{GHZ}^{(n-1)}$	& $\theta$ \\
\toprule
\endhead
$\mathcal{B}^{(2)}$	& $\sqrt{2}\approx 1.4142$	& {}	  				& $\frac{3\pi}{4}\approx 2.3562$  \\ 
$\mathcal{B}^{(3)}$	& $1.5$			 	& $\frac{3}{2\sqrt{2}}\approx 1.0607$	& $\frac{2\pi}{3}\approx 2.0944$  \\ 
$\mathcal{B}^{(4)}$	& $1.5539$		 	& $ 1.0359$ 			  	& $ 1.9786$  \\ 
$\mathcal{B}^{(5)}$	& $1.5926$		 	& $ 1.0249$ 			  	& $ 1.9106$  \\ 
$\mathcal{B}^{(6)}$	& $1.6224$		 	& $ 1.0187$ 			  	& $ 1.8650$  \\ 
$\mathcal{B}^{(7)}$	& $1.6464$		 	& $ 1.0148$ 			  	& $ 1.8318$  \\ \toprule
\label{table:table}
\end{longtable}
For a given number of parties $n$, the corresponding relation in~\eqref{GHZvalue_simplified} can be numerically optimized w.r.t. $\theta$ and the limits become 
\begin{align}
\lim\limits_{n\to\infty}^{} g^{(n)}_{\text{GHZ}} = 2 \quad \quad \text{and} \quad \quad \lim\limits_{n\to\infty}^{} \theta^{(n)} = \frac{\pi}{2},
\end{align}
which is visualized in Fig.~\ref{fig:bellvalue}.

\begin{figure}[ht!]
\centering
\includegraphics[width=0.475\textwidth]{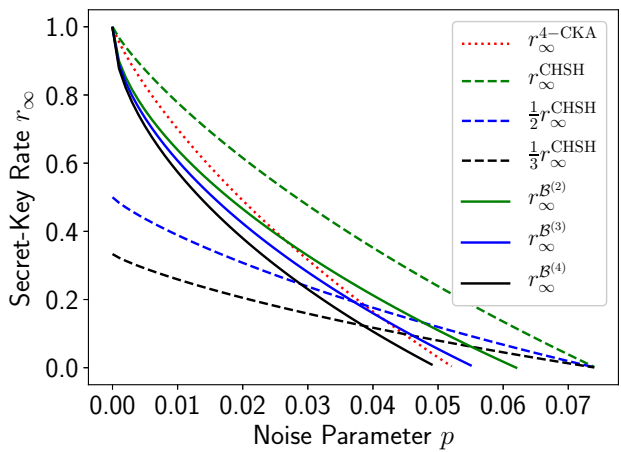}
\caption{Asymptotic DI secret-key rates according to Eq.~\eqref{DIkeyrate2} in dependence of the noise parameter $p$ (solid lines) for $n\in\{2,3,4\}$. 
In bottleneck networks and for low noise, the multipartite DIQKD protocol outperforms multiple bipartite DIQKD protocols, Eq.~\eqref{Pironio}, 
(dashed lines). The dotted line corresponds to the analytical bounds of Ref.~\cite{DICKAfixed}, Eq.~$(4)$ for $n=4$ in the same implementation. 
In terms of key rates calculated via SDP, however, our Bell inequality leads to better results compared to 
the Parity-CHSH inequality (not shown in this Figure), an advantage that increases with 
the noise parameter $p$. For example for $n=3$ and $p\in\{3,4,5\}\%$, key rates based on $\mathcal{B}^{(3)}$ are larger by approximately 
$\{1.2,3.6,16.8\}\%$.}
\label{fig:noisemodels}
\end{figure}

From Table~\ref{table:table}, we notice that the Bell value $g_\text{GHZ}^{(n)}$ for $n\in\{3,4\}$ coincides with the Tsirelson bound in 
Eq.~\eqref{tsirelson34}. 
Due to the symmetry and construction of the Bell inequality, we conjecture that this holds for general $n$. If this is true, finding the Tsirelson 
bound to our Bell inequality boils down to a simple numerical optimization over the parameter $\theta$ in Eq.~\eqref{GHZvalue_simplified}. 
To conclude this discussion, consider the Bell inequality for $n=3$ parties. States of 
the form $\rho=\rho_{AB^{(2)}}\otimes\rho_{B^{(3)}}$ do not allow to exceed the Tsirelson bound for $n=2$ parties, which one can verify -- either 
analytically or via the NPA hierarchy -- by taking all classical deterministic strategies for Bob$^{(3)}$ into account. Thus, $\sqrt{2}$ is a 
Svetlichny bound~\cite{svetlichny} which can certify genuine tripartite entanglement. 
Likewise, one observes that states of the form $\rho=\rho_A\otimes\rho_{B^{(2)}B^{(3)}}$ cannot violate the classical bound. 
Beyond the tripartite case, we have numerical indication for analogous statements concerning biseparable splits, cf. Outlook.\\
\emph{Bounding Eves guessing probability.---} 
Finally, we want to apply our Bell inequality~\eqref{BellIneqClass} for DIQKD. As preparation, we briefly describe how to obtain a 
lower bound on the DI conference key rates. 
We focus on asymptotic secret-key rates and assume that quantum devices behave identically and independently in each round (i.i.d.). 
Let $\mathcal{G}^{(n)}$ denote the Bell operator corresponding to our Bell inequality~\eqref{BellIneqClass}, i.e., 
$\mathcal{B}^{(n)} = \tr{\mathcal{G}^{(n)}\rho_{A\boldsymbol{B}}}$, where $\rho_{A\boldsymbol{B}}\coloneqq\rho_{AB^{(2)}\dots B^{(n)}}$ represents 
the quantum state shared among all parties. Let Alice use measurement input $x = 0$ for raw key generation 
and define $\boldsymbol{B}_{\boldsymbol{y}}\coloneqq (B_{y^{(2)}}^{(2)},\dots,B_{y^{(n)}}^{(n)})$. 
Eve's guessing probability $P_\text{g}(\boldsymbol{a}\vert\mathcal{E})$ about Alice's $A_0$-measurement results $\boldsymbol{a}$ conditioned on her 
information $\mathcal{E}$ can be upper bounded by a function $f$ of the observed Bell violation $g_\text{obs}^{(n)}$, i.e., 
$P_\text{g}(\boldsymbol{a}\vert\mathcal{E})\leqslant f(g_\text{obs}^{(n)})$. For fixed $g_\text{obs}^{(n)}$, it amounts to the solution of the 
SDP~\cite{Masanes, npa, wittek}
\begin{alignat}{3}\label{SDP}
\max\limits_{\rho_{A\boldsymbol{B}}, A_{x},\boldsymbol{B}_{\boldsymbol{y}}}  \,\,\,	& \tr{A_0\rho_{A\boldsymbol{B}}} &&  \\
\text{subject to: }	& \text{tr}\big(\mathcal{G}^{(n)} \rho_{A\boldsymbol{B}}\big)= \gobs^{(n)}&&.  \nonumber
\end{alignat}
For classical-quantum states $\rho_{A\mathcal{E}}$, the guessing probability is connected to the quantum min-entropy via 
$H_\text{min}\left(\bm{a}\vert\mathcal{E}\right)=-\log_2\pguess(\bm{a}\vert\mathcal{E})$~\cite{operationalmeaning}, 
from which we obtain a lower bound on the DI asymptotic secret-key rate, $r_{\infty,n}^\text{SDP} \geqslant -\log_2f\left(\gobs\right) - h\left(Q\right)$, 
where $h(p)\coloneqq-p\log_2(p) - (1-p)\log_2(1-p)$ and $Q$ denote the binary entropy and the quantum bit error rate (QBER), respectively. 
The noisiest channel determines the QBER~\cite{NQKD}, hence
\begin{align}
 Q = \max\limits_{j\in[n]}^{}\left(Q_{AB^{(j)}}\right), \label{QBER}
\end{align}
where $Q_{AB^{(j)}}$ is the QBER between Alice and Bob$^{(j)}$. 
The bound established by the SDP~\eqref{SDP} is valid against the most general attacks the eavesdropper can perform~\cite{Masanes} but they are 
in general rather loose. Recent development promises improvement in this regard~\cite{rene}.\\
\emph{Application: DI conference key agreement.---} Here, we present achievable DI secret-key rates for $n \in \{2,3,4\}$ parties with a DIQKD protocol 
similar to the one in Ref.~\cite{DICKA}. In the honest implementation, the quantum state distributed in each round of the protocol is the 
$n$-GHZ state. To minimize the error-correction information, all parties measure $\sigma_z$ in key generation rounds. 
To test for Bell-inequality violation, the parties choose observables as proposed in Eq.~\eqref{optimalchoice} that lead to a 
maximum violation. The protocol is aborted if the Bell inequality~\eqref{BellIneqClass} 
is not violated. For a realistic scenario, we assume local depolarizing noise, that corrupts each qubit subsystem $\rho_i$ according to
\begin{align}
\mathcal{D}_\text{dep}\left(\rho_i\right) = (1-p)\rho_i + \frac{p}{2}\mathds{1}_2,   \label{dep}
\end{align}
where $p\in[0,1]$ denotes the noise parameter. In this scenario, the marginal probability distribution of Alice's $A_0$ measurement is uniform, i.e., 
$\left\langle A_0\right\rangle = 0$. Since we consider binary outcomes, 
we can lower bound the Von Neumann entropy in terms of the guessing probability via 
$H(\bm{a}\vert\mathcal{E}) \geqslant 2\left(1-\pguess(\bm{a}\vert\mathcal{E})\right)$~\cite{rene0,rene}, which in turn yields
\begin{align}
 r_{\infty}^{\mathcal{B}^{(n)}} \geqslant 2\left(1-\pguess(\bm{a}\vert\mathcal{E})\right) - h\left(Q\right). \label{DIkeyrate2}
\end{align}
Figure~\ref{fig:noisemodels}  
displays the lower bound on the asymptotic DI secret-key rate, Eq.~\eqref{DIkeyrate2}, as a function of the parameter $p$ of 
the noise model in Eq.~\eqref{dep}. To put 
these key rates into perspective, we consider the same comparison as in Ref.~\cite{DICKAfixed}, where the conference key rates are compared 
with multiple bipartite key rates, described by~\cite{scaraniDI}
\begin{align}
r_\infty^\text{CHSH}\!\geqslant 1-h\left(Q\right) - h\!\left(\!\frac{1+\sqrt{S^2\slash4-1}}{2}\right) \label{Pironio},
\end{align}
where $S$ denotes the violation of the CHSH inequality. 
For illustration, we consider the Bell state $\ket{\phi^+}\propto\ket{00}+\ket{11}$ under the noise model in Eq.~\eqref{dep}, which connects $S$ with $Q$ 
according to $S=2\sqrt{2}\left(1-2Q\right)$. The QBER $Q$ as defined in Eq.~\eqref{QBER} is related to the noise parameter via 
$Q=p\left(1-p\slash2\right)$ for all $n$. 
Under the assumption that Alice cannot perform the bipartite QKD protocols with every Bob simultaneously, which can be the case in bottleneck networks, 
cf. Ref.~\cite{NQKD}, the bipartite key rates get a prefactor of ${(n-1)}^{-1}$.\\
As mentioned, the bounds on the guessing probability in terms of SDPs are often too pessimistic. Therefore, we cannot beat 
the analytical results of Ref.~\cite{DICKAfixed}. In direct comparison via the SDP, however, 
our Bell inequality leads to slightly better conference key rates than the Parity-CHSH inequality, see caption of Fig.~\ref{fig:noisemodels}.\\
\emph{Conclusion and Outlook.---} In this manuscript, we introduced a novel family of genuine multipartite Bell 
inequalities, that is specifically tailored to the $n$-GHZ state, while maintaining the possibility to maximally violate it with $\sigma_z$ measurements. 
As argued, an application is to use this Bell inequality for a Bell test in a DIQKD protocol, because there highly correlated measurement results 
and maximal violation are required at the same time. We established the classical bounds of this Bell inequality and suggested measurements that 
lead to the maximal Bell value, given the $n$-GHZ state is measured. Finally, we calculated via semidefinite programming conference key 
rates based on the violation of our Bell inequality and discussed its robustness against depolarizing noise.\\
For future work, a more thorough study of our Bell inequality~\eqref{BellIneqClass} is desirable. A starting point is to clarify the role of 
partially entangled states and the existence of associated intermediate bounds in our Bell inequality, similar to the MABK case~\cite{wernerMABK}. 
We conjecture that the maximum Bell value $\mathcal{B}^{(n)}$ for $n$ parties with biseparable states where at most $k-1$ Bobs are entangled with Alice, 
is determined by the maximum Bell value $\mathcal{B}^{(k)}$ for $k$ parties. 
In this case a Bell value larger then $\mathcal{B}^{(k)}$ is a DI witness 
for entanglement of at least $k+1$ parties, one of them being Alice. 
An important goal would be to find an analytical bound on the Von Neumann entropy in terms of the violation of our 
Bell inequality~\eqref{BellIneqClass}. As we provided a nontrivial genuinely multipartite generalization of the CHSH inequality -- in a similar spirit as 
the $n$-GHZ state represents a multipartite generalization of the Bell state -- 
we hope that our contribution paves the way for further insight into multipartite quantum communication.

\begin{acknowledgments}
The authors acknowledge support from the Federal Ministry of Education and Research BMBF (Project Q.Link.X and HQS) and from ML$4$Q Excellence 
Cluster of DFG. We thank Reinhard Werner, Gl\'{a}ucia Murta, and Lucas Tendick for helpful discussions.
\end{acknowledgments}
\appendix
\begin{widetext}
\section{Supplemental Material}
We split the Suppl. Mat. into four parts. First, we prove the classical upper and lower bounds of our Bell inequality. 
Afterwards, we elaborate on the construction of the Bell inequality and discuss optimal measurements to achieve 
a maximum Bell value with the $n$-GHZ state. Finally we state the DIQKD protocol for completeness. 
We recall our Bell inequality for convenience:
\begin{align}\label{BellIneqClassApp}
-\left(2^{n-1}-1\right)&\leqslant\left\langle A_1 \bigotimes\limits_{j=2}^{n}B_+^{(j)}\right\rangle 
- \delta_{\lfloor\frac{n}{2}\rfloor,\frac{n}{2}} \left\langle A_0 \bigotimes\limits_{j=2}^{n}B_-^{(j)}\right\rangle \\ \nonumber 
&-\sum\limits_{k=1}^{\lfloor\frac{n-1}{2}\rfloor}\left[
\left\langle 
A_0\otimes \sum\limits_{\boldsymbol{\alpha}^{(n)}_{2k-1}\in\mathcal{S}^{(n)}_{2k-1}}\bigotimes\limits_{j=1}^{2k-1} B_-^{\left(\alpha^{(n)}_{2k-1,j}\right)}
\right\rangle + \left\langle 
\sum\limits_{\boldsymbol{\alpha}^{(n)}_{2k}\in\mathcal{S}^{(n)}_{2k}}\bigotimes\limits_{j=1}^{2k} B_-^{\left(\alpha^{(n)}_{2k,j}\right)}
\right\rangle 
\right] \leqslant 1.
\end{align}

\section{Proof of the Theorem}
The maximal and minimal classical value is achieved for deterministic strategies. To establish the classical bounds, we thus consider the variables 
$A_x$ and $B_{y^{(j)}}^{(j)}$ for $x,y^{(j)}\in\{0,1\}$, $j \in \{2,\dots,n\}$ to take on values from the set $\{\pm 1\}$ and denote with the vector 
$\big(\boldsymbol{A},\boldsymbol{B}^{(j)}\big)$ a strategy from the set that contains every possible combination of $\pm1$ as components for this 
$2n$-dimensional vector. We also define 
\begin{align}
\tilde{\mathcal{B}}^{(n)}\coloneqq  - \delta_{\lfloor\frac{n}{2}\rfloor,\frac{n}{2}}  A_0 \prod\limits_{j=2}^{n}B_-^{(j)}
- \sum\limits_{k=1}^{\lfloor\frac{n-1}{2}\rfloor}\left[
A_0 \sum\limits_{\boldsymbol{\alpha}^{(n)}_{2k-1}\in\mathcal{S}^{(n)}_{2k-1}}\prod\limits_{j=1}^{2k-1} B_-^{\left(\alpha^{(n)}_{2k-1,j}\right)}
+ 
\sum\limits_{\boldsymbol{\alpha}^{(n)}_{2k}\in\mathcal{S}^{(n)}_{2k}}\prod\limits_{j=1}^{2k} B_-^{\left(\alpha^{(n)}_{2k,j}\right)}
\right], \label{Btilde}
\end{align}
such that we can write $\mathcal{B}^{(n)}= A_1\prod_jB_+^{(j)} + \tilde{\mathcal{B}}^{(n)}$ for the classical Bell value. 
We make the important observation, that any 
strategy $\left(\boldsymbol{A},\boldsymbol{B}^{(j)}\right)$ that leads to $A_1\prod_jB_+^{(j)}\neq0$, eliminates the value of $\tilde{\mathcal{B}}^{(n)}$ 
as this requires that $B_+^{(j)}\neq0$ (and thus $B_-^{(j)}=0$) for all $j\in[n]$. Therefore, we can maximize and minimize the expressions 
$A_1\prod_jB_+^{(j)}$ and $\tilde{\mathcal{B}}^{(n)}$ 
independently. This distinction into cases allows us, to map the strategies for the maximization (minimization) of $\tilde{\mathcal{B}}^{(n)}$ from 
$\left(\boldsymbol{A},\boldsymbol{B}^{(j)}\right) \in \{\pm1\}^{2n}$ to $\big(A_0, \boldsymbol{B}_-^{(j)}\big)$ with $A_0\in\{\pm1\}$ and 
$B_-^{(j)}=\frac{1}{2}\big(B_0^{(j)}-B_1^{(j)}\big)\in\{\pm1,0\}$. 
For the proof we require three important properties of the binomial coefficients:
\begin{subequations}
\label{binomprops}
\begin{align}
\sum\limits_{l=0}^{n}\binom{n}{l} &= 2^{n} &\text{(Normalization)}, \label{norm} \\
\binom{n}{l} &= \binom{n-1}{l} + \binom{n-1}{l-1} &\text{(Pascal triangle)}, \label{pascal}\\
\binom{n}{l} &= \sum\limits_{j=0}^l \binom{m}{j}\binom{n-m}{l-j} &\text{(Chu-Vandermonde identity)} \label{chuvan}.
\end{align}
\end{subequations}
Note, that we make use of the conventions $0! = 1$ and $\binom{n}{l} = 0$ $\forall l>n, l<0$. 
We divide the proof into two parts, one for the lower and one for the upper bound.\\
\noindent \emph{(i) Lower bound.} 
To establish the lower classical bound, note that the minimization of $A_1\prod_jB_+^{(j)}$ leads only to the value of $-1$. A minimization of 
$\tilde{\mathcal{B}}^{(n)}$, however, is given by the choice $B_-^{(j)}=+1$ for all $j$ and $A_0=+1$, as this turns every contribution in 
Eq.~\eqref{Btilde} negative, in detail 
\begin{align}
\tilde{\mathcal{B}}^{(n)} = -\delta_{\lfloor\frac{n}{2}\rfloor,\frac{n}{2}} - \sum\limits_{k=1}^{\lfloor\frac{n-1}{2}\rfloor}
\left[\binom{n-1}{2k-1}+\binom{n-1}{2k}\right] =
-\delta_{\lfloor\frac{n}{2}\rfloor,\frac{n}{2}} - \sum\limits_{k=1}^{2\lfloor\frac{n-1}{2}\rfloor}\binom{n-1}{k}, 
\end{align}
where we used the cardinality $\#\mathcal{S}_l^{(n)}=\binom{n-1}{l}$. Via the normalization condition, Eq.~\eqref{norm}, 
the expression above simplifies for both $n$ odd and even to $-\big(2^{n-1} -1 \big)$, as claimed.\\
\noindent \emph{(ii) Upper bound.}
A maximization of $A_1\prod_jB+^{(j)}$ leads to the value of $1$, but a priori it is not clear that this is indeed the maximum possible 
$\mathcal{B}^{(n)}$-value. We start by counting all possible strategies for $\tilde{\mathcal{B}}^{(n)}$ and categorize them, such that we can calculate its 
value by a distinction of cases. There are $2\times3^{n-1}$ different possibilities to choose a strategy $\big(A_0,\boldsymbol{B}_-^{(j)}\big)$, however, 
we notice that the expression $\tilde{\mathcal{B}}^{(n)}$ in Eq.~\eqref{Btilde} is invariant under permutation of Bobs, i.e., we only need to calculate the 
$\tilde{\mathcal{B}}^{(n)}$-value for a subset of strategies $\big(A_0,\boldsymbol{B}_-^{(j)}\big)$, that cannot be converted into each other by permutation 
of Bobs. This reduces the number of different deterministic strategies to only $n(n+1)$. 
As a final remark before we work through the different strategies note that the amount of nonzero values for the variables $B_-^{(j)}$ determines 
which summands give a nontrivial contribution to $\tilde{\mathcal{B}}^{(n)}$. 
To be more specific, let $q_\pm$ denote the amount of $\pm1$-values in the strategy $\big(A_0,\boldsymbol{B}_-^{(j)}\big)$, and let 
$q_++q_-  \eqqcolon q \leqslant n-1$ be the amount of nonzero $B_-^{(j)}$-values. Due to the permutational invariance of $\tilde{\mathcal{B}}^{(n)}$ we 
order without loss of generality the strategy $\big(A_0, \boldsymbol{B}_-^{(j)}\big)$ such that $B_-^{(j)}=0$ for all $j>q+1$. Then, 
every product in Eq.~\eqref{Btilde} associated to a label 
$\boldsymbol{\alpha}_l^{(n)}\in\mathcal{S}^{(n)}_l\setminus\mathcal{S}^{(q+1)}_l$ vanishes, as it contains at least one Bob$^{(j)}$ with 
$B_-^{(j)}=0$. This converts the sum over the set $\mathcal{S}^{(n)}_l$ into a sum over the set $\mathcal{S}^{(q+1)}_l$ of cardinality 
$\binom{q}{l}$. The expression $A_0\prod_{j=2}^nB_-^{(j)}$ always vanishes for $q<n-1$.\\ 
\emph{(a) $q\leqslant n-1, q = q_\pm$.} For these cases, $B_-^{(j)}=B_-^{(k)}$ holds for all $j,k \in \{2,\dots,q+1\}$. Applying this strategy, yields
\begin{align}
\tilde{\mathcal{B}}^{(n)}&=-\delta_{\lfloor \frac{n}{2}\rfloor, \frac{n}{2}}\delta_{n-1,q} A_0 (\pm 1)^{n-1} 
-\sum\limits_{k=1}^{\lfloor\frac{n-1}{2}\rfloor} \left[A_0\sum\limits_{\alpha_{2k-1}^{(q+1)} \in \mathcal{S}^{(q+1)}_{2k-1}}(\pm 1)^{2k-1} 
+\sum\limits_{\alpha_{2k}^{(q+1)} \in \mathcal{S}^{(q+1)}_{2k}} (\pm 1)^{2k}
\right] \nonumber \\ 
&=-\delta_{\lfloor \frac{n}{2}\rfloor, \frac{n}{2}}\delta_{n-1,q} A_0 (\pm 1)^{n-1} 
-\sum\limits_{k=1}^{\lfloor\frac{n-1}{2}\rfloor} \left[\pm A_0 \binom{q}{2k-1} + \binom{q}{2k}
\right].
\label{qequalsn}
\end{align}
To proceed, let $n$ be an odd integer, hence $\delta_{\lfloor\frac{n}{2}\rfloor,\frac{n}{2}}=0$. Then, the best Alice can do is to choose her 
variable $A_0 \in \{\pm1\}$ such that the sum is minimized, because of the global 
minus sign in Eq.~\eqref{qequalsn}. Exploiting identity~\eqref{pascal}, leads to 
\begin{align}
\tilde{\mathcal{B}}^{(n)} =  - \sum\limits_{k=1}^{\frac{n-1}{2}} \left[ - \binom{q-1}{2k-2} + \binom{q-1}{2k} \right],
\end{align}
where the only nonvanishing term is $\binom{q-1}{0}$ and thus results in $\tilde{\mathcal{B}}^{(n)} = 1$. For $n$ even, we can make a 
similar argument. Choosing the 
value for $A_0$ that maximizes the total expression leads us to 
\begin{align}
\tilde{\mathcal{B}}^{(n)} =  \delta_{n-1,q} +\binom{q-1}{0} - \binom{q-1}{n-2} = 
\delta_{n-1,q} + 1 - \delta_{n-1,q} = 1. 
\end{align}
\emph{(b)} $q_{+} + q_- = q \leqslant n-1, q_\pm\geqslant1$. For the remaining cases, at least one variable $B_-^{(j)}$ is $+1$ and at least one is $-1$. 
From Eq.~\eqref{Btilde} we obtain with this strategy
\begin{align}
\tilde{\mathcal{B}}^{(n)}&=
(-1)^{q_-+1}\delta_{\lfloor\frac{n}{2}\rfloor,\frac{n}{2}}\,\delta_{n-1,q}\,A_0 
-\sum\limits_{k=1}^{\lfloor\frac{n-1}{2}\rfloor}\left[A_0\sum\limits_{r=0}^{2k-1}(-1)^r\binom{q_+}{2k-1-r}\binom{q_-}{r} 
+ \sum\limits_{r=0}^{2k}(-1)^r\binom{q_+}{2k-r}\binom{q_-}{r}
\right]. \label{alternatingChuVan}
\end{align}
Recall, that in the case where all Bobs have the same value, we have $\#\mathcal{S}^{(q+1)}_{l}$ combinations to attribute 
the value $\pm1$ to all $l$ out of $q$ Bobs. Here, the sum still has $\binom{q}{l}$ many terms, but some multiply to $+1$, while others to $-1$, 
depending on how many elements are drawn from $q_-$. To correctly count the numbers of combinations leading to the sign $\pm1$, we 
use the Chu-Vandermonde identity~\eqref{chuvan}. The idea here is to divide the total amount of options $q$ into two subsets $q_+$ and $q_-$, and 
then count all possible combinations to draw elements from these 
subsets. But due to the negativity of elements from the set $q_-$, we need to include a negative sign for $\binom{q_-}{r}$ if $r$ is odd. 
Important is, that due to the alternating sign, almost all terms in Eq.~\eqref{alternatingChuVan} cancel each other. In fact, the following two relations 
hold 
\begin{subequations}
\begin{align}
\sum\limits_{k=1}^{\lfloor\frac{n-1}{2}\rfloor} \left[\sum\limits_{r=0}^{2k-1}(-1)^r\binom{q_+}{2k-1-r}\binom{q_-}{r} 
\right] &= (-1)^{q_-+1}\delta_{\lfloor \frac{n}{2}\rfloor, \frac{n}{2}}\delta_{n-1, q} &&\text{and} \label{claim1}\\ 
\sum\limits_{k=1}^{\lfloor\frac{n-1}{2}\rfloor}
\left[\sum\limits_{r=0}^{2k}(-1)^r\binom{q_+}{2k-r}\binom{q_-}{r}
\right] &= -1 &&\forall\,\,\, n\in\mathbb{N},\,\, q_\pm\geqslant1,\,\, q_++q_-\leqslant n-1. \label{claim2} 
\end{align}
\end{subequations}
Showing the validity of these relations concludes the prove, as inserting them into  
Eq.~\eqref{alternatingChuVan} leads to the maximum of $\tilde{\mathcal{B}}^{(n)} = 1$. To prove Eq.~\eqref{claim1} we order the 
left-hand side of it by positive and negative contributions
\begin{subequations}
\begin{align}
\sum\limits_{k=1}^{\lfloor\frac{n-1}{2}\rfloor} \left[\sum\limits_{r=0}^{2k-1}(-1)^r\binom{q_+}{2k-1-r}\binom{q_-}{r}\right] =  \label{sub21a}
&\sum\limits_{k=1}^{\lfloor\frac{n-1}{2}\rfloor} \left[\sum\limits_{r=0}^{k-1}\binom{q_+}{2k-1-2r}\binom{q_-}{2r}\right]  \\-
&\sum\limits_{k=1}^{\lfloor\frac{n-1}{2}\rfloor} \left[\sum\limits_{r=0}^{k-1}\binom{q_+}{2k-1-(2r+1)}\binom{q_-}{2r+1}\right]. \label{sub21b}
\end{align}
\end{subequations}
The idea is to use the Pascal triangle relation~\eqref{pascal}, to eliminate the problems that arise due to the alternating sign. Via 
Eq.~\eqref{pascal} we thus split the right-hand side of Eq.~\eqref{sub21a} into the following two expressions:
\begin{subequations}
\begin{align}
\sum\limits_{k=1}^{\lfloor\frac{n-1}{2}\rfloor}\sum\limits_{r=0}^{k-1}\binom{q_--1}{2r}\left[\binom{q_+-1}{2k-1-2r}+\binom{q_+-1}{2(k-1)-2r}\right]
&= \!\!\sum\limits_{x=0}^{2\lfloor\frac{n-1}{2}\rfloor - 1} \sum\limits_{r=0}^{\lfloor\frac{x}{2}\rfloor}\binom{q_--1}{2r}\!\binom{q_+-1}{x-2r} , \label{22a}\\ 
\sum\limits_{k=2}^{\lfloor\frac{n-1}{2}\rfloor}\sum\limits_{r=1}^{k-1}\binom{q_--1}{2r-1}\left[\binom{q_+-1}{2(k-1)-(2r-1)}+
\binom{q_+-1}{2(k-1)-1-(2r-1)}\right]
&= \!\!\sum\limits_{x=1}^{2\lfloor\frac{n-1}{2}\rfloor - 2} \sum\limits_{r=1}^{\lfloor\frac{x+1}{2}\rfloor}
\!\!\!\!\binom{q_--1}{2r-1}\!\binom{q_+-1}{x-(2r-1)}, \label{22b}
\end{align}
\end{subequations}
where we introduced a new index of summation $x$ to simplify both expressions. We dropped the contributions from $k=1$ and $r=0$ in Eq.~\eqref{22b}, 
as they vanish anyway. To proceed, we add the right-hand sides of Eqs.~\eqref{22a} and~\eqref{22b}. All integers from $r=0$ up to $r=x$, for 
all $x \in\{0,\dots,2\lfloor\frac{n-1}{2}\rfloor - 2\}$ appear in this sum. 
Therefore, the right-hand side of Eq.~\eqref{sub21a} is given by 
\begin{align}
\sum\limits_{k=1}^{\lfloor\frac{n-1}{2}\rfloor} \left[\sum\limits_{r=0}^{k-1}\binom{q_+}{2k-1-2r}\binom{q_-}{2r}\right] 
= \sum\limits_{x=0}^{2\lfloor\frac{n-1}{2}\rfloor - 2}\sum\limits_{y=0}^x \binom{q_--1}{y}\binom{q_+-1}{x-y} + 
\sum\limits_{r=0}^{\lfloor\frac{n-1}{2}\rfloor -1}\binom{q_--1}{2r}\binom{q_+-1}{2\lfloor\frac{n-1}{2}\rfloor - 1 -2r}, \label{23}
\end{align}
where we used $\lfloor \lfloor\frac{n-1}{2}\rfloor- \frac{1}{2}\rfloor = \lfloor\frac{n-1}{2}\rfloor-1$. 
To simplify Eq.~\eqref{23}, note that the second sum only yields a nontrivial contribution, if $2r\leqslant q_--1$ 
and $2r\geqslant2\lfloor\frac{n-1}{2}\rfloor-q+q_-$, 
which is only possible if $q\geqslant2\lfloor\frac{n-1}{2}\rfloor + 1$. As we additionally have the constraint $q\leqslant n-1$, we require $q=n-1$ 
and $n$ needs to be an even 
integer. In this case, the only nonvanishing term in the second sum in Eq.~\eqref{23} is a single expression equal to $+1$, corresponding to 
$r=\frac{q_--1}{2}$, which can only be a valid integer if $q_-$ is odd. Beyond this, 
we use the Chu-Vandermonde identity~\eqref{chuvan} to simplify the first expression of the right-hand side in Eq.~\eqref{23} and obtain
\begin{align}
\sum\limits_{k=1}^{\lfloor\frac{n-1}{2}\rfloor} \left[\sum\limits_{r=0}^{k-1}\binom{q_+}{2k-1-2r}\binom{q_-}{2r}\right] 
= \sum\limits_{x=0}^{2\lfloor\frac{n-1}{2}\rfloor - 2}\binom{q-2}{x} + \delta_{n-1,q}\delta_{\lfloor\frac{n}{2}\rfloor,\frac{n}{2}}
\delta_{\lfloor\frac{q_--1}{2}\rfloor,\frac{q_--1}{2}}. \label{24}
\end{align}
The same procedure can be applied to the right-hand side of Eq.~\eqref{sub21b}. Ultimately, it leads to 
\begin{align}
\sum\limits_{k=1}^{\lfloor\frac{n-1}{2}\rfloor} \left[\sum\limits_{r=0}^{k-1}\binom{q_+}{2k-1-(2r+1)}\binom{q_-}{2r+1}\right] 
= \sum\limits_{x=0}^{2\lfloor\frac{n-1}{2}\rfloor - 2}\binom{q-2}{x} + \delta_{n-1,q}\delta_{\lfloor\frac{n}{2}\rfloor,\frac{n}{2}}
\delta_{\lfloor\frac{q_-}{2}\rfloor,\frac{q_-}{2}}, \label{25}
\end{align}
where the additional contribution is now only obtained if $q_-$ is an even integer. The difference between Eqs.~\eqref{24} and~\eqref{25} represents the 
left-hand side of Eq.~\eqref{sub21a}. We thus obtain 
\begin{align}
\sum\limits_{k=1}^{\lfloor\frac{n-1}{2}\rfloor} \left[\sum\limits_{r=0}^{2k-1}(-1)^r\binom{q_+}{2k-1-r}\binom{q_-}{r}\right] 
=\delta_{n-1,q}\delta_{\lfloor\frac{n}{2}\rfloor,\frac{n}{2}}\left(\delta_{\lfloor\frac{q_--1}{2}\rfloor,\frac{q_--1}{2}}-
\delta_{\lfloor\frac{q_-}{2}\rfloor,\frac{q_-}{2}}\right) = 
(-1)^{q_-+1}\delta_{n-1,q}\delta_{\lfloor\frac{n}{2}\rfloor,\frac{n}{2}}, \label{26}
\end{align}
which proves identity~\eqref{claim1}. 
Essentially the same approach now leads to the prove of relation~\eqref{claim2}. Only minor and straightforward adjustments for the index of 
summations are needed, which then leads to 
\begin{align}
\sum\limits_{k=1}^{\lfloor\frac{n-1}{2}\rfloor}\left[\sum\limits_{r=0}^{2k}(-1)^r\binom{q_+}{2k-r}\binom{q_-}{r}
\right] = -\binom{q-2}{0} + \binom{q-2}{2\lfloor\frac{n-1}{2}\rfloor - 1} -\delta_{n-1,q}\delta_{\lfloor\frac{n}{2}\rfloor,\frac{n}{2}} = -1,
\end{align}
because the second binomial coefficient is $+1$ if $n$ is even and $q=n-1$, and $0$ otherwise. This concludes the proof. \hfill $\blacksquare$

\section{On the Construction of the Bell Inequality} 
The Bell inequality~\eqref{BellIneqClassApp} is constructed around two central restrictions 
we impose on the Bell setting. First, we want to achieve a large Bell value if the quantum resource is given by an $n$-GHZ state and second, that 
this Bell value is achievable if Alice measures $A_0=\sigma_z$. As the Bell inequality is tested for violation in a DIQKD protocol, these restrictions 
are clearly motivated by Theorem $1$ of Ref.~\cite{NQKD}, which states that maximum correlation among all $n$ parties with a GHZ state requires all 
parties to measure $\sigma_z$. 
We set the stage by discussing known multipartite Bell inequalities and introducing some notation. 
A priori, it is not clear, how to devise a useful Bell inequality, that is particularly well suited for the $n$-GHZ state. 
The MABK inequality~\cite{mermin, ardehali, belinskii} for instance allows a maximum violation by the $n$-GHZ state, as discussed in Ref.~\cite{werner}. 
For DIQKD however, the MABK inequality is not suitable because the very structure of it prohibits to simultaneously achieve perfectly 
correlated measurement results among all parties and sufficiently high Bell-inequality violation, see Ref.~\cite{comment} for details. Also 
most recently, Ref.~\cite{tailoredBell} introduces a Bell inequality which is tailored to be maximally violated by an $n$-GHZ state of any local 
dimension $d$. However, at least for $d=2$ and $m=2$ measurement settings, this inequality suffers from the same drawbacks as the MABK inequality. 
Imposing the additional constraint on the Bell setting, that Alice should in principle be able to measure $A_0=\sigma_z$ without compromising the 
possibility to violate the Bell inequality has led us to our inequality~\eqref{BellIneqClassApp}. Another Bell inequality which 
embraces this idea, is the Parity-CHSH inequality~\cite{DICKAfixed} 
\begin{align}
\mathcal{B}_\text{Parity}^{(n)} \coloneqq A_1 \otimes \frac{B_0^{(2)}+B_1^{(2)}}{2}\bigotimes\limits_{j=3}^nB^{(j)}
-A_0\otimes\frac{B_0^{(2)}-B_1^{(2)}}{2}
\leqslant 1\leqslant\sqrt{2} , \label{parity}
\end{align}
where each Bob$^{(j)}$ for $j\geqslant3$ only has one observable. In fact, the Parity-CHSH inequality can be reproduced from our Bell 
inequality~\eqref{BellIneqClassApp}, by choosing $B_0^{(j)}=B_1^{(j)}$ for all $j\geqslant3$ and therefore $B_-^{(j)} = 0$ 
and $B_+^{(j)}=B_0^{(j)} \eqqcolon B^{(j)}$.

We briefly recall the notation we already introduced in Ref.~\cite{comment}, as it is crucial for the construction of the Bell 
inequality~\eqref{BellIneqClassApp}. 
Let $\mathbb{F}_2=\{0,1\}$ denote the finite field with two elements, which allows us to define the 
vector space $\mathbb{F}_2^n$ of bit strings of length $n$. Let further $\mathcal{P}_n$ denote the $n$-qubit Pauli group. We define 
the stabilizer group
\begin{align}
\mathcal{S} \coloneqq \Big\{S \in \mathcal{P}_n \,\,\Big\vert\,\, S \ket{\text{GHZ}_n} = \ket{\text{GHZ}_n} \Big\}
\end{align}
of the $n$-GHZ state $\chi_n=\ket{\text{GHZ}_n}\!\bra{\text{GHZ}_n}$. 
The group $\mathcal{S}$ is generated by the $n$ independent operators
\begin{subequations}
\label{generators}
\begin{align}
G_1&\coloneqq \sigma_x^{\otimes n}, \quad \text{and for all } j \in [n]: \\ 
G_j&\coloneqq \bigotimes\limits_{i=1}^{j-2}\mathds{1}_2^{(i)}\otimes \sigma_z^{(j-1)}\otimes\sigma_z^{(j)}\otimes\bigotimes
\limits_{i=j+1}^{n}\mathds{1}_2^{(i)},
\end{align}
\end{subequations}
where the superscript denotes the corresponding subsystems. In general, the projector of any stabilizer state can be written as the normalized sum 
of all of its stabilizer operators~\cite{stabilizer0,stabilizer00}. We obtain for $\chi_n$ with $\bm{s}\coloneqq(s_1,\dots,s_n) \in \mathbb{F}_2^n$ 
the representation:
\begin{align}
\chi_n = \frac{1}{2^n}\sum\limits_{\bm{s}\in\mathbb{F}_2^n}\left(\sigma_x^{s_1}\right)^{\otimes n} &\big(\sigma_z^{s_2} \otimes  
\sigma_z^{s_2+s_3} \otimes \dots\otimes \sigma_z^{s_{n-1}+s_n} \otimes \sigma_z^{s_n}\big). \label{generalGHZ}
\end{align}
The sum in Eq.~\eqref{generalGHZ} consists of $2^n$ individual terms, where $2^{n-1}$ of them contain only Pauli $\sigma_z$ and identity operators 
(namely those with $s_1=0$), while the other $2^{n-1}$ ones consists of only Pauli $\sigma_x$ and $\sigma_y$ operators. The 
\emph{weight} of such operators is given by the number of nontrivial Pauli matrices it contains. For $s_1=1$, the operators always have full weight, 
while for $s_1=0$ the weight of the operators is always an 
even number, but all possible combinations (with respect to the subsystems) of all even numbers $2k\leqslant n$ of $\sigma_z$ occur. 
For the construction of our Bell inequality, we pursue a strategy which matches the restrictions we initially imposed on the Bell setting. To obtain 
a large quantum value with the $n$-GHZ state, the idea is to gain a contribution from as many operators as possible from the representation 
in Eq.~\eqref{generalGHZ}. To quantify this, recall that Pauli matrices are traceless and that their product is given by
\begin{align}
\sigma_j \sigma_k = \delta_{j,k}\mathds{1}_2 + i \sum\limits_{l=1}^{3}\epsilon_{jkl}\sigma_l , \label{paulirelation}
\end{align}
where $\delta_{j,k}$ and $\epsilon_{jkl}$ denote the Kronecker delta and the Levi-Civita tensor, respectively. As we require $A_0=\sigma_z$ and because of 
relation~\eqref{paulirelation} the expression 
\begin{align}
\text{tr}\bigg[A_0  \bigotimes\limits_{j\in\mathcal{I}}\big(B_0^{(j)} - B_1^{(j)}\big)\sum\limits_{\bm{s}\in\mathbb{F}_2^n, s_1=1}
\big(\sigma_x\sigma_z^{s_2} \otimes \dots \otimes \sigma_x\sigma_z^{s_n}\big)\bigg] = 0\label{counterpart}
\end{align}
always vanishes, for any index subset $\mathcal{I}\subseteq\{2,\dots,n\}$, for all $\boldsymbol{s}$ with $s_1=1$ and for all dichotomic observables 
$B_i^{(j)}$. 
The counterpart of 
expression~\eqref{counterpart} for $s_1=0$ however, is nonvanishing if the observables have an even weight. 
The same argument can be done for the corresponding expression without an observable of Alice. As all possible combinations occur in the $n$-GHZ state, we also 
include all possible combinations of observables with respect to the parties for expectation values in our Bell inequality. 
This explains the term 
\begin{align}
-\sum\limits_{k=1}^{\lfloor\frac{n-1}{2}\rfloor}\Bigg[
\left\langle 
A_0\otimes \sum\limits_{\boldsymbol{\alpha}^{(n)}_{2k-1}\in\mathcal{S}^{(n)}_{2k-1}}\bigotimes\limits_{j=1}^{2k-1} B_-^{\left(\alpha^{(n)}_{2k-1,j}\right)}
\right\rangle + \left\langle 
\sum\limits_{\boldsymbol{\alpha}^{(n)}_{2k}\in\mathcal{S}^{(n)}_{2k}}\bigotimes\limits_{j=1}^{2k} B_-^{\left(\alpha^{(n)}_{2k,j}\right)}
\right\rangle 
\Bigg]
\end{align}
in our Bell inequality. The expression $-\delta_{\lfloor\frac{n}{2}\rfloor,\frac{n}{2}} \big\langle A_0 \bigotimes_{j=2}^{n}B_-^{(j)}\big\rangle$ is included 
due to a fundamental difference between the odd- and even-numbered $n$-GHZ state. For $n$ even, the operator $\sigma_z^{\otimes n}$ occurs in the GHZ 
state representation in Eq.~\eqref{generalGHZ}, while for $n$ odd, this is not the case. 
Finally, since operators with $s_1=1$ have full weight, we include one additional expectation value in the Bell inequality that contains 
observables of all parties, 
hence the first term in our Bell inequality.
\section{Optimal Measurements and Properties of the Bell Inequality} 
As our main goal was to establish a useful Bell inequality for multipartite device-independent quantum key distribution (DIQKD), our focus is not the complete 
characterization of our Bell inequality. For completeness, however, we want to address some properties, in particular 
we suggest measurement observables for all parties that lead to a maximum Bell value if the $n$-GHZ state is measured, because this is relevant for QKD. 
Further properties which could be worth investigating are, if it is possible to analytically derive the Tsirelson bounds~\cite{tsirelson}, 
if the Bell inequalities 
constitute facets of the classical polytope~\cite{locpoly}, or if there exist intermediate bounds for separable states with respect to different splits of 
parties, as it is the case for the MABK inequality~\cite{wernerMABK}. For $n=3$ we discovered that $\mathcal{B}^{(3)}$ is in fact a facet inequality, as 
one can show 
with the methods presented in Ref.~\cite{facet}. As already mentioned in the main article, we conjecture that there exist intermediate bounds.\\
To motivate the optimal choices for the observables given the GHZ state $\chi_n$ is measured, recall that a general qubit observable can be parametrized as 
\begin{align}
B_i^{(j)} = \cos\big(\varphi_i^{(j)}\big)\sin\big(\theta_i^{(j)}\big)\sigma_x + 
\sin\big(\varphi_i^{(j)}\big)\sin\big(\theta_i^{(j)}\big)\sigma_y + \cos\big(\theta_i^{(j)}\big)\sigma_z, \label{generalQubit}
\end{align}
and analogously for $A_1$. Note that $B_0^{(j)},B_1^{(j)}$ always appear as 
$B_-^{(j)}\propto B_0^{(j)}-B_1^{(j)}$  in our Bell inequality, if paired with $A_0$ or if no observable of Alice is included. 
To maximize the corresponding expectation values, it is best to eliminate the contribution of all $B_-^{(j)}$ in $\sigma_x$ and $\sigma_y$ 
direction, as this part vanishes anyway due to the structure of the GHZ state in Eq.~\eqref{generalGHZ}. 
This translates to $\varphi_0^{(j)} = \varphi_1^{(j)}$ for all $j\in[n]$, as a necessary condition to guarantee $B_-\propto \sigma_z$.
Likewise, the expression $B_+^{(j)}\propto B_0^{(j)}+B_1^{(j)}$ appears only in combination with $A_1$. Because all operators with $s_1=1$ 
in Eq.~\eqref{generalGHZ} have full weight, we might as well take that $A_1$ and all $B_+^{(j)}$ expressions 
have no contribution in $\sigma_z$ direction, to gain a large contribution to the Bell value from $\left\langle A_1\bigotimes_{j=2}^n B_+^{(j)}\right\rangle$. 
Due to $\cos(\alpha)=-\cos(\pi\pm\alpha)$, we extract $\theta_1^{(j)} = \pi \pm \theta_0^{(j)}$ for all $j\in[n]$ from the representation~\eqref{generalQubit}, 
as a necessary condition to eliminate the $\sigma_z$ contribution of $B_+^{(j)}$. 
Beyond that, we note that $\sin(\pi\pm\alpha)=\mp\sin(\alpha)$. Together with $\varphi_0^{(j)} = \varphi_1^{(j)}$, the 
choice $\theta_1^{(j)} = \pi + \theta_0^{(j)}$ eliminates $B_+^{(j)}$, which is why we use 
$\theta_1^{(j)} = \pi - \theta_0^{(j)}$ in the following. Finally, we numerically find that for a given choice of $A_1$, the actual value of the 
azimuthal angle 
$\varphi_0^{(j)}, \varphi_1^{(j)}$ is irrelevant for maximizing the Bell value, as long as they are equal for each Bob. Therefore, we set 
$\varphi_0^{(j)}=\varphi_1^{(j)}=0$ for all $j\in[n]$ and $\varphi_{A_1}=0$. Furthermore, the polar angles $\theta_0^{(j)},\theta_1^{(j)}$ can be 
chosen the same for 
every Bob, without compromising the possibility to achieve the maximum Bell value. We therefore set $\theta_0^{(j)}=\theta$ and 
$\theta_1^{(j)}=\pi - \theta$ for all $j\in[n]$. In total, the maximum Bell value $\mathcal{B}^{(n)}$ given an $n$-GHZ state is measured, can be achieved with
\begin{align}
A_0=\sigma_z, \quad A_1=\sigma_x,\quad B_0^{(j)} = \sin\left(\theta\right)\sigma_x + \cos\left(\theta\right)\sigma_z, \quad 
B_1^{(j)} = \sin\left(\theta\right)\sigma_x - \cos\left(\theta\right)\sigma_z\quad \forall\,\,j\in[n], \label{QvalApp}
\end{align}
where the optimal value of the polar angle $\theta$ depends on the number of parties $n$. This choice allows a straightforward calculation of 
the Bell value with the $n$-GHZ state
\begin{align}
 g^{(n)}_{\text{GHZ}}= \left[\sin\left(\theta\right)^{n-1} - \delta_{\lfloor\frac{n}{2}\rfloor,\frac{n}{2}}\cos\left(\theta\right)^{n-1}\right]
 -\sum\limits_{k=1}^{\lfloor\frac{n-1}{2}\rfloor}\cos\left(\theta\right)^{2k-1}\left[\binom{n-1}{2k-1}+\cos\left(\theta\right)\binom{n-1}{2k}\right],
\end{align}
which can be simplified to 
\begin{subequations}
\label{GHZvalue_simplifiedApp}
\begin{align}
g^{(n)}_{\text{GHZ}}&= 1- \left(1+\cos\left(\theta\right)\right)^{n-1}+ \sin\left(\theta\right)^{n-1} &\text{for }n\text{ odd}, \\ 
g^{(n)}_{\text{GHZ}}&= 1- \left(1+\cos\left(\theta\right)\right)^{n-1}+ 
\frac{\cot\left(\theta\slash2\right)\sin\left(\theta\right)^n}{1+\cos\left(\theta\right)}
&\text{for }n\text{ even}.
\end{align}
\end{subequations}
For given $n$, the corresponding relation~\eqref{GHZvalue_simplifiedApp} can be numerically optimized for $\theta$ and the limits become 
\begin{subequations}
\begin{align}
\lim\limits_{n\to\infty}^{} g^{(n)}_{\text{GHZ}} = 2 \quad\quad \text{and} \quad \quad \lim\limits_{n\to\infty}^{} \theta^{(n)} = \frac{\pi}{2}. 
\end{align}
\end{subequations}
\section{Multipartite DIQKD Protocol} 
Finally, we want to state the DIQKD protocol. Alice has two measurement inputs 
$x\in\{0,1\}$ implementing the measurement of a dichotomic observable $A_x$. 
Each Bob$^{(j)}$ has three inputs $y^{(j)} \in \{0,1,2\}$, with dichotomic observables $B_{y^{(j)}}^{(j)}$. The protocol includes the 
following steps, see also~\cite{NQKD,DICKA}:
\begin{enumerate}[label=(\roman*)]
\item In every round of the protocol, the parties do:\\
\emph{State preparation -} Alice produces and distributes a multipartite state $\rho_{A\boldsymbol{B}}$. Since we assume an i.i.d. implementation, the source 
generates the same state in every round.\\
\emph{Measurement -} There are two types of measurement rounds, key generation (type-$0$) and parameter estimation (type-$1$) measurement rounds. 
For type $0$, the parties choose the inputs $(x,\boldsymbol{y}) = (0,2,\dots,2)$, and for type $1$ they choose their inputs $x,y^{(j)} \in\{0,1\}$ uniformly 
at random. The parties use a preshared random key to agree on the type of measurement round.
\item \emph{Parameter estimation -} 
The parties publicly communicate the list of bases and outcomes for type-$1$ rounds and an equal amount of measurement outputs 
for type-$0$ rounds. The publicly announced data from type $1$ is used to estimate the Bell value $\gobs^{(n)}$ of 
inequality~\eqref{BellIneqClassApp}, whereas the announced 
type-$0$ data is used to estimate the quantum bit error rate $Q$, which quantifies the asymptotic error-correction information. 
\item \emph{Classical postprocessing -} Similar to the device-dependent multipartite QKD protocol~\cite{NQKD}, an error-correction and privacy-amplification 
protocol is performed. 
\end{enumerate} 
If the parties verify, that their data violates our Bell inequality~\eqref{BellIneqClassApp}, they commence the error correction. 
The solution of the SDP in the article then upper bounds Eve's guessing probability. If $\gobs^{(n)} \leqslant g_\text{cl}^{(n)\downarrow}$ they abort the 
protocol. 
\end{widetext}
\bibliography{Belln22TimoHolz}
\end{document}